# Pump-probe study of plasma dynamics in gas-filled photonic crystal fiber using counter-propagating solitons

M. I. Suresh,* F. Köttig, J. R. Koehler, F. Tani, and P. St.J. Russell

*Max Planck Institute for the Science of Light, Staudtstrasse 2, 91058 Erlangen, Germany*
*Corresponding author: mallika-irene.suresh@mpl.mpg.de

**Abstract:** We present a pump-probe technique for monitoring ultrafast polarizability changes. In particular, we use it to measure the plasma density created at the temporal focus of a self-compressing higher-order pump soliton in gas-filled hollow-core photonic crystal fiber. This is done by monitoring the wavelength of the dispersive wave emission from a counter-propagating probe soliton. By varying the relative delay between pump and probe, the plasma density distribution along the fiber can be mapped out. Compared to the recently introduced interferometric side-probing for monitoring the plasma density, our new technique is relatively immune to instabilities caused by air turbulence and mechanical vibration. The results of two experiments on argon- and krypton-filled fiber are presented, and compared to numerical simulations. The technique provides an important new tool for probing photoionization in many different gases and gas mixtures as well as ultrafast changes in dispersion in many other contexts.

I. Introduction

Studies of photoionization by strong optical fields have led to a number of interesting developments in physics, ranging from fundamental studies of ionization and subsequent plasma evolution [1–3] to applications such as high harmonic generation (HHG) [1] and the formation of plasma channels that can be used in particle acceleration and for the generation of coherent X-rays [4]. While these initial studies focused on low-density gases, recent research has explored the full nonlinear polarization of gases at atmospheric and higher pressures [5], as well as more exotic effects such as Kramers-Henneberger quasi-bound states [6], and avalanche ionisation by collisions of excited atoms in the absence of a laser pulse [7–9]. Understanding the dynamics of ionised gases at high pressures is also of great importance for HHG in the water window, a spectral region where phase-matching generally requires helium pressures of several atmospheres [10,11].

While these effects have mostly been investigated in free space, gas-filled hollow-core photonic crystal fiber (HC-PCF) has over the past few years emerged as a highly suitable platform for precise studies of laser-plasma interactions [12]. Avoiding the usual limitations imposed by beam diffraction in free-space experiments, HC-PCF efficiently guides few-femtosecond pulses at intensities sufficient for ionization. These features have made possible the observation of a plasma-driven soliton self-frequency blue-shift [13,14], a mechanism for emission of mid-infrared dispersive waves [15] and plasma-induced soliton fission [16]. Compared to conventional capillaries, the much smaller core in HC-PCF makes it possible to reach ionizing intensities at pulse energies 1000 times lower, permitting the pump laser repetition rate to be scaled to MHz rates without increasing the average power to unmanageable levels. At higher repetition rates, however, ionization-induced changes in refractive index can have lifetimes long enough to disturb the propagation of subsequent pulses [17], an effect that has also been seen in femtosecond enhancement cavities [18]. Recently we have studied such effects in gas-filled HC-PCF over a timespan of tens of µs with a temporal resolution of ~1 ns, both interferometrically [19] and using a prism-coupling technique [20].

Here we report a pump-probe technique for monitoring the photoionization-induced free electron density by self-compressing solitons along a gas-filled HC-PCF, including its temporal evolution up to a few ns after the ionization event. The key feature of the technique is the use of a counter-propagating probe pulse that emits dispersive wave (DW) radiation in the ultraviolet at its temporal focus. In the presence of a plasma, the effective refractive index (modal index) of the gas-filled fiber decreases, which changes the phase-matching condition for DW emission, shifting the DW to a higher frequency. This non-interferometric technique is very robust against noise caused by air turbulence and mechanical vibrations. The counter-propagating configuration also allows easy separation of pump and probe signals, while eliminating cross-phase modulation between them. The experimental results are compared with numerical simulations, showing good agreement.

## II.   Theory and experiment

Dispersive waves, the light shed from a perturbed higher-order soliton to linear waves [21], have been widely exploited to generate wavelength-tunable radiation from the VUV to mid-IR [12, 15, 22]. They are emitted when there is phase-matching between the soliton and a dispersive (i.e., linear) wave. Neglecting fiber anti-crossings, the propagation constant of the LP$_{pm}$-like mode in a HC-PCF with core radius $a$ is given to a good approximation by [12]:

$$\beta(\omega) = (\omega/c)n_{pm} = (\omega/c)\sqrt{n_{gas}^2 - u_{pm}^2 c^2/(\omega a)^2}, \tag{1}$$

where $n_{pm}$ is the modal index, $\omega$ is the optical angular frequency, $c$ the speed of light in vacuum, $n_{gas}$ the pressure-dependent refractive index of the gas, and $u_{pm}$ the $m$-th zero of the $p$-th order Bessel function of the first kind.

The dephasing rate between a soliton and a DW at frequency $\omega$ can be written as:

$$\Delta\beta(\omega) = \beta(\omega) - \beta_{sol}(\omega) = \beta(\omega) - (\beta_0 + \beta_1[\omega - \omega_0] + \gamma P_c/2), \tag{2}$$

where $\beta_0$ is the propagation constant and $\beta_1$ the inverse group velocity at the central frequency $\omega_0$ of the soliton, $P_c$ the peak power of the compressed soliton and $\gamma$ the pressure-dependent Kerr coefficient. Dispersive wave emission occurs when $\Delta\beta(\omega) = 0$, a condition that will be affected by any dispersive change in $n_{gas}$, resulting in a frequency shift of the DW and thus providing a means of probing local changes in dispersion along the fiber.

A higher-order soliton undergoes temporal self-compression as it propagates along a gas-filled PCF. Under appropriate conditions it can reach a peak intensity high enough to ionize the gas, creating a free electron density $N_e$ that causes refractive index changes. On a femtosecond timescale these are purely caused by the presence of free electrons, whereas recombination-driven heating of the gas within typically several nanoseconds would result in thermal and acoustic effects lasting a few hundred microseconds [19, 23]. The initial refractive index of a partially ionized gas is given by [24]:

$$n_{gas} = \sqrt{n_0^2 - \frac{\bar{N}_e e^2}{\varepsilon_0 m_e \omega^2}} \approx n_0\left(1 - \frac{\bar{N}_e e^2}{2\varepsilon_0 m_e \omega^2 n_0^2}\right), \tag{1}$$

where $n_0$ is the index of the neutral gas, $m_e$ and $e$ are electronic mass and charge, $\varepsilon_0$ is the vacuum permittivity and $\bar{N}_e$ is the free electron density, assumed constant across the core. Equations (1) and (3) predict that $\bar{N}_e \sim 3\times10^{15}$ cm$^{-3}$ will cause $n_{pm}$ to fall by $8\times10^{-7}$ at 800 nm and $5\times10^{-8}$ at 200 nm in a fiber with core diameter 34 µm. Depending on the peak power of a probe soliton, such a variation in dispersion can cause the DW to shift, for example by 0.5 nm with a 2 µJ probe pulse, which is easily detectable by a conventional spectrometer.

Fig. 1(a) plots $\Delta\beta(\omega)$ for a HC-PCF with 34 µm core diameter filled with 4.1 bar of argon, showing that the DW blue-shift increases with increasing free electron density. For $\bar{N}_e = 10^{17}$ cm$^{-3}$ (a modest value), the DW wavelength shifts by 15 nm (inset of Fig. 1(a)) and for $\bar{N}_e = 10^{18}$ cm$^{-3}$, it shifts by 80 nm. In reality, however, the plasma generated by the self-compressing soliton is localized within a small region near the center of the core and varies strongly along the fiber length. As a result the measured DW wavelength shift is smaller than that predicted by one-dimensional solutions of $\Delta\beta(\omega) = 0$. To account for this, we used an effective value of the spatially dependent plasma density as $\bar{N}_e$ in the numerical simulations, as described in detail below.

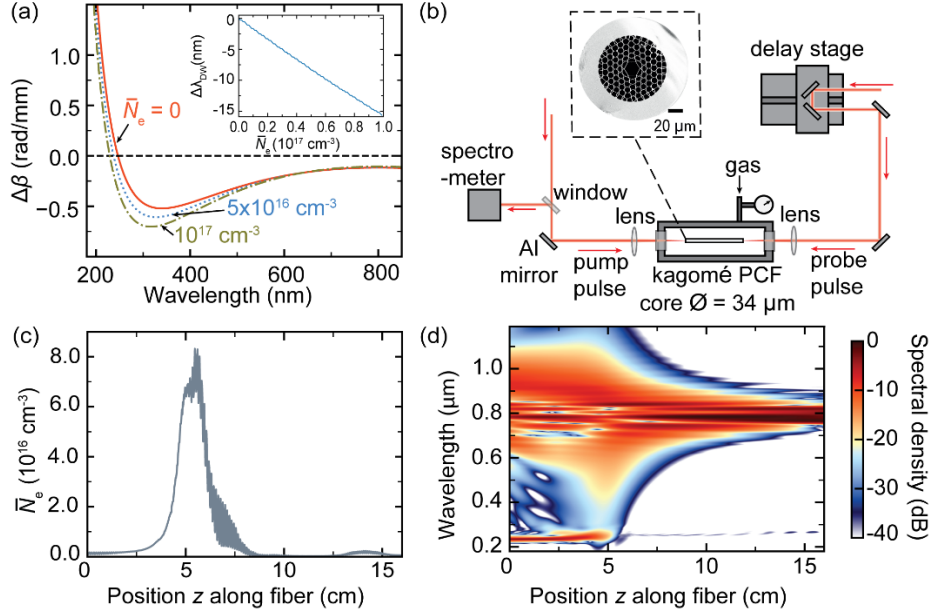

**Figure 1.** (a) DW dephasing rate for different values of plasma density, showing the shift in the phase-matching wavelength $\lambda_{DW}$, calculated for 1.4 µJ pulses propagating in 4.1 bar Ar. Inset: shift in DW wavelength $\Delta\lambda_{DW}$ as a function of $\bar{N}_e$. (b) Experimental set-up: Pump and probe pulses counter-propagate in gas-filled kagomé HC-PCF, core diameter 34 µm. Inset: SEM of the fiber microstructure. (c) Simulated plasma density $\bar{N}_e(z)$ along a 16 cm length of this HC-PCF filled with 4.1 bar of Ar and pumped by a forward-propagating 6 µJ pulse. (d) Simulated spectral evolution of a 1.4 µJ probe pulse propagating backwards along the same gas-filled fiber as in (c), with the pump blocked.

Fig. 1(b) sketches the experimental set-up. Pulses from a Ti:sapphire laser amplifier (central wavelength 800 nm, 25 fs FWHM pulse duration, 1 kHz repetition rate) were divided at a 67:33 fused-silica beam splitter and delivered to two arms. In each arm a combination of half-wave plate and thin-film polarizer was used to control the pulse energies. Uncoated calcium fluoride plano-convex lenses with 15 cm focal length were used to launch the pulses into opposite ends of a kagomé HC-PCF with a core diameter of 34 µm (inset of Fig. 1(b)) and an estimated propagation loss of 2 dB/m at 800 nm. The fiber was kept in a pressurized gas cell with magnesium fluoride windows to allow in- and out-coupling of light. The probe light emerging from the fiber is reflected into a spectrometer at a fused-silica window mounted at 45°.

In the probe arm, a retroreflector mounted on a linear translation stage was used to vary the relative delay between pump and probe in computer-controlled steps of 13.3 ps over a maximum delay of 2.6 ns (limited by the travel range of the stage). The delay $\tau$ is defined such that the pulses collide at $z \sim \tau c/2$ where $c$ is the vacuum speed of light, i.e., for $\tau = 0$ they meet at $z = 0$. Prior to mounting the fiber, we calibrated the system to meet this condition (to an accuracy of ±1 mm) by placing a prism mirror at the pump input end, directing both counter-propagating pulses into a beta barium borate crystal and maximizing the second-harmonic intensity to determine the temporal overlap. In addition, with the fiber in place, we noticed a sharp drop (a few percent) in the probe power when it encountered the plasma at $\tau = 0$, allowing the calibration to be checked.

Two different experiments were performed. In the first, 6.6 µJ pump and 2.0 µJ probe pulses were launched into a 16-cm-long HC-PCF filled with 4.1 bar of argon (zero-dispersion wavelength ~515 nm). In the second, a 14.3 cm length of fiber was filled with 2.1 bar of krypton (zero-dispersion wavelength ~525 nm) and the energies reduced to 6.4 µJ for the pump and 1.8 µJ for the probe so as to ensure that in each experiment the DW of the probe is emitted close to the maximum of the plasma density profile generated by the pump. In both cases the dispersion was anomalous at the laser wavelength, the soliton order of the probe pulses being $N \sim 5$, while for the pump pulses $N \sim 9$ in argon and $N \sim 10$ in krypton, ensuring that the conditions for temporal self-compression were fulfilled.

Numerical simulations were carried out using the unidirectional full-field nonlinear wave equation [25] with $\chi^{(3)}$ values from [26]. To account for imperfect launching and optical losses, the energies used in the simulations were slightly lower than in the experiments: 6 µJ pump and 1.4 µJ probe in argon and 6.2 µJ pump and 1.5 µJ probe in krypton. The phase and amplitude profiles of the pulses at the fiber input were estimated from measurements made (using spectral phase interferometry for direct electric-field reconstruction—SPIDER) just before the delay stage and used in the simulations, taking account of the dispersion introduced by the subsequent optical components. Photoionization was included in the simulations using Perelomov, Popov, Terent'ev rates, modified with the Ammosov, Delone, Krainov coefficients [27], while disregarding effects of plasma recombination.

The numerical simulations accounted for the effects of both axial propagation and transverse field variations, enabling us to extract the radial and axial variation in plasma density $N_e(r, z)$ generated by the pump, and thus the on-axis plasma density $N_e(0, z)$. Finite element modeling (FEM) was used to numerically calculate the effective index $n_{01}(z)$ of the $LP_{01}$-like mode in the presence of $N_e(r, z)$, assuming the core to be a simple capillary. Equations (1) and (3) were then found to yield a good fit to $n_{01}(z)$ for a uniform plasma density $\bar{N}_e(z) \sim 0.15 \times N_e(0, z)$. This is the quantity plotted in Fig. 1(c) for the argon case, with the pump pulse entering the fiber from the left. In contrast, simulations of the probe pulse dynamics were one-dimensional, being based for simplicity on the nonlinear effective area. The calculated spectral evolution of a 1.4 µJ probe pulse in a fiber filled with 4.1 bar of argon (with the pump blocked) is shown in Fig. 1(d). No plasma forms under these conditions. To simulate the effect of ionization on the probe dynamics, we used $\bar{N}_e$ to calculate the refractive index of the ionized gas (Eq. (3)). A simplified schematic of this procedure is given as a flowchart in the Supplemental Material [28]. The effect of varying the pump-probe delay was explored by increasing the region of non-zero plasma density traversed by the probe pulse in steps, starting at the pump input end.

### III.    Results and discussion

The measured spectra of the probe-driven DWs are plotted against pump-probe delay $\tau$ in Fig. 2(a) for the first (argon) experiment and in Fig. 2(c) for the second (krypton) experiment. The vertical dashed lines indicate the calculated delay at which the pulses meet at the probe input end: $\tau = (2 \times L)/c$ which is $\tau \approx 1.07$ ns for the HC-PCF filled with argon with $L = 0.16$ m and $\tau \approx 0.95$ ns for the case with krypton where $L = 0.143$ m. In each plot, the white curve represents the spectral centroid $\lambda_{DW}$ of the DW, calculated over a wavelength range of 50 nm. The standard deviation of $\lambda_{DW}$ over three consecutive measurements was always less than 1 nm.

From Fig. 2(a) for the experiment in argon (Fig. 2(c) for krypton), we see that before the pump pulse enters the fiber ($\tau < 0$) the probe pulse emits a DW centered at 231 nm in neutral argon and 244 nm in neutral krypton. With the onset of photoionization at delays $\tau > 0$, the probe DW shifts to 224 nm in argon and 237 nm in krypton. We note from Fig. 1(d) that the probe DW is always emitted at $z \sim 5$ cm (i.e. at $\tau = (2 \times z)/c \sim 333$ ps), so that its wavelength will shift only if the pump pulse has already ionized the gas at that location. This shift slowly recovers as the free electrons recombine with their parent ions. The value of $\bar{N}_e$ was calibrated to the measured changes in $\lambda_{DW}$ by comparison with numerical simulations.

The agreement with numerical simulations (Fig. 2(b)) is good for the argon experiment (Fig. 2(a)), while for the krypton experiment, the wavelength of the probe DW is slightly longer in the experiments (Fig. 2(c)) than in the simulations (Fig. 2(d)), even in the absence of a plasma (i.e., for $\tau < 0$). This could be caused by a combination of a slightly lower $\chi^{(3)}$ than reported in [26], variations in core diameter along the fiber, and excitation of higher-order modes at the launch end as a result of self-focusing effects in the gas cell (these are stronger in Kr). In both experiments, the peak power of the probe pulse at the position of DW emission (~1 GW in each case) is below the critical power for self-focusing (2.8 GW in argon and 2.0 GW in krypton).

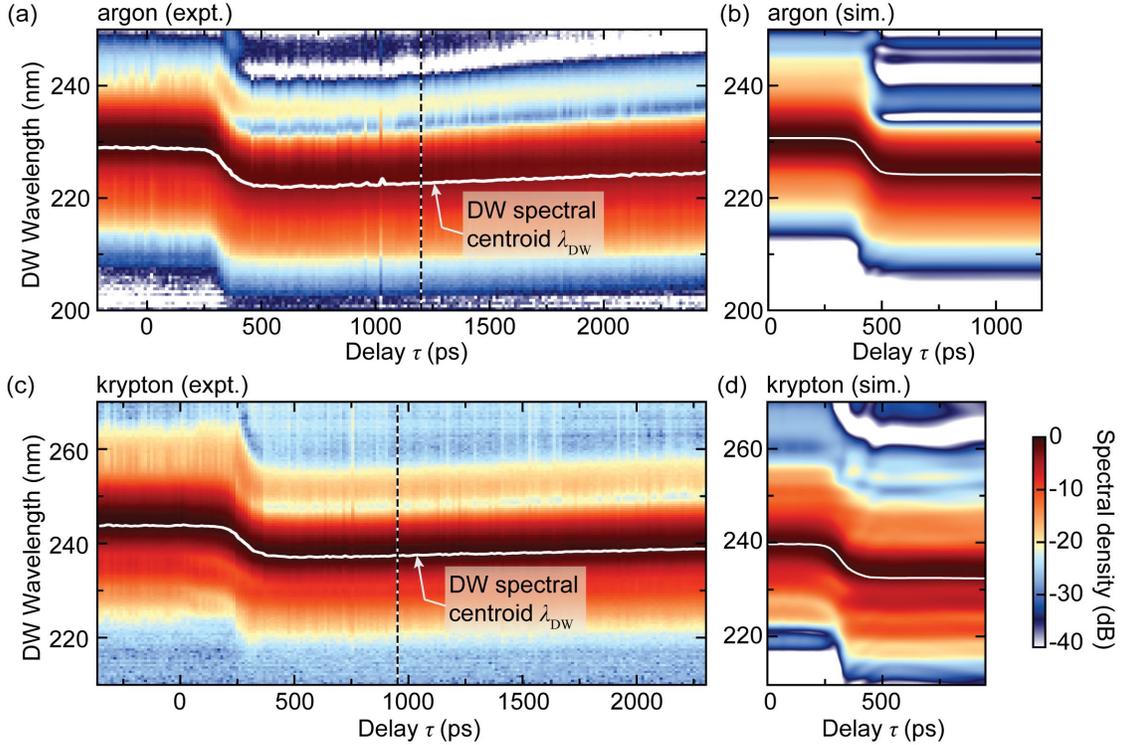

**Figure 2.** Spectral density of the probe DW and its spectral centroid $\lambda_{DW}$ (white curve) plotted against the delay between pump and probe pulses $\tau$. The black dashed lines indicate the delay for which the pulses meet at the probe end of the fiber, given by $\tau = 2L/c$. (a) Argon experiment with $L = 0.16$ m. (b) Numerical simulation of argon experiment. (c) Krypton experiment, $L = 0.143$ m. (d) Numerical simulation of krypton experiment.

The best fit between the numerical and experimental values of $\lambda_{DW}$ was found for $\bar{N}_e(z) \sim 0.09 \times N_e(0, z)$, not $0.15 \times N_e(0, z)$ as predicted from FEM calculations (See the Supplemental Material [28]). We attribute this discrepancy to short-timescale plasma density changes and slight reductions in probe energy during propagation through the plasma, which are not accounted for in the modeling. Hence, we note from Fig. 2(a) that in argon, the probe measures $\bar{N}_e = 4 \times 10^{16}$ cm$^{-3}$ at $\tau \sim 600$ ps. This drops to $\bar{N}_e = 2 \times 10^{16}$ cm$^{-3}$ at $\tau \sim 2450$ ps as the electrons recombine. In krypton, as shown in Fig. 2(c), this calculated drop in plasma density is less pronounced: from $\bar{N}_e = 5.2 \times 10^{16}$ cm$^{-3}$ at $\tau \sim 540$ ps to $\bar{N}_e = 3.6 \times 10^{16}$ cm$^{-3}$ at $\tau \sim 2364$ ps. Note that the simulations in Fig. 2(b, d) do not account for plasma recombination, so do not reproduce the slow return of the DW wavelength to its initial value at $\tau < 0$.

For $\tau < 250$ ps, the measurements are sensitive to the build-up of the plasma along the fiber within the ~1-2 cm over which the probe DW is emitted. As $\tau$ is increased, the temporal overlap between DW emission by the probe and the plasma increases, causing $\lambda_{DW}$ to gradually shift to shorter wavelength. The derivative of this wavelength shift, $d\lambda_{DW}/d\tau$, is approximately proportional to $\bar{N}_e(z = \tau c/2)$ within the DW emission region, which is plotted in Fig. 3 for the two cases.

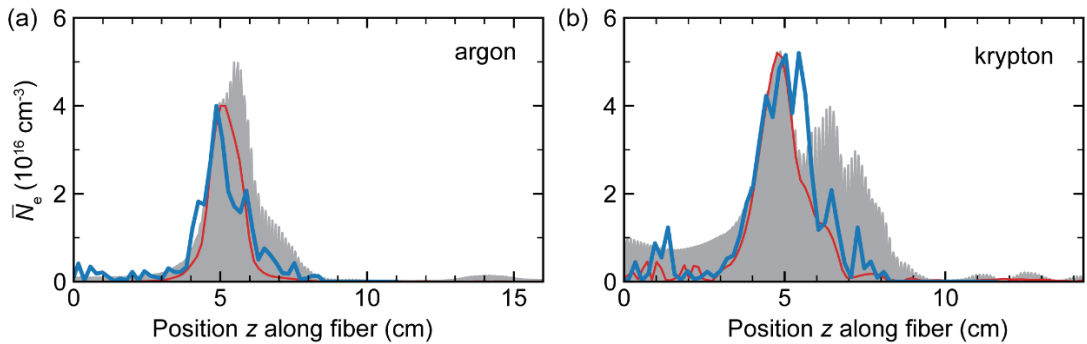

**Figure 3.** Profile of the effective plasma density along the HC-PCF, reconstructed from the experimental data (thick blue line) and the numerical experiments (red thin line) for (a) argon and (b) krypton. $\bar{N}_e(z)$ obtained from the simulated pump pulse propagation is indicated by gray shaded areas.

Reconstructions of the plasma profile are plotted in Figs. 3(a, b) for the argon and krypton cases, the blue curves representing the experimental data and the red curves the numerical simulations. Both agree well with the simulated pump plasma profiles (gray shaded areas) within the probe DW emission region, which was carefully aligned with the temporal focus of the pump pulse. We note that the plasma profile can be measured by scanning the region of DW emission along the fiber, e.g. by chirping the probe pulse ([29]) or changing the probe power. The numerical values of plasma density and its distribution along the fiber are noted to be consistent with previous results in which different measurement techniques were used [19, 20].

## IV. Conclusions

The plasma density created at the temporal focus of a self-compressing higher-order soliton in gas-filled HC-PCF can be elegantly probed by monitoring the wavelength of DW emission from a counter-propagating soliton, which carries lower energy and does not generate significant plasma itself. By varying the relative delay between pump and probe pulses, the plasma density distribution along the fiber can be mapped out within the length of the DW emission. Compared to interferometric side-probing for monitoring plasma density [19], this approach is less subject to instabilities caused by air turbulence and mechanical vibrations. The results of two experiments on argon- and krypton-filled HC-PCF are in good agreement with numerical simulations. The length of fiber probed can be extended by shifting emission region of the DW in the fiber by varying the chirp of the probe pulse. The technique provides an important new tool for probing photoionization in many different gases and gas mixtures in hollow-core fibers and capillaries. Systems in which DW emission may be useful include microresonators and silicon waveguides, where it would enable similar changes in dispersion caused by the creation of free carriers to be monitored. Other phase-matched processes, such as conical emission, could also be used to probe dispersion changes in free-space filamentation.

## References

[1] F. Calegari, G. Sansone, S. Stagira, C. Vozzi, and M. Nisoli, "Advances in attosecond science," *J. Phys. B. At. Mol. Opt. Phys.* **49**, 062001 (2016).

[2] L. Young, D. A. Arms, E. M. Dufresne, R. W. Dunford, D. L. Ederer, C. Höhr, *et al.*, "X-ray microprobe of orbital alignment in strong-field ionized atoms," *Phys. Rev. Lett.* **97**, 083601 (2006).

[3] S. Tzortzakis, B. Prade, M. Franco, and A. Mysyrowicz, "Time-evolution of the plasma channel at the trail of a self-guided IR femtosecond laser pulse in air," *Opt. Commun.* **181**, 123–127 (2000).

[4] C. G. Durfee, J. Lynch, and H. M. Milchberg, "Development of a plasma waveguide for high-intensity laser pulses," *Phys. Rev. E* **51**, 2368–2389 (1995).

[5] J. K. Wahlstrand, S. Zahedpour, A. Bahl, M. Kolesik, and H. M. Milchberg, "Bound-electron nonlinearity beyond the ionization threshold," *Phys. Rev. Lett.* **120**, 183901 (2018).

[6] M. Matthews, F. Morales, A. Patas, A. Lindinger, J. Gateau, N. Berti, *et al.*, "Amplification of intense light fields by nearly free electrons," *Nat. Phys.* **14**, 695–700 (2018).

[7] X. Gao, G. Patwardhan, S. Schrauth, D. Zhu, T. Popmintchev, H. C. Kapteyn, *et al.*, "Picosecond ionization dynamics in femtosecond filaments at high pressures," *Phys. Rev. A* **95**, 013412 (2017).

[8] D. A. Romanov, X. Gao, A. L. Gaeta, and R. J. Levis, "Intrapulse impact processes in dense-gas femtosecond laser filamentation," *Phys. Rev. A* **97**, 063411 (2018).

[9] E. M. Wright, S. W. Koch, M. Kolesik, and J. V. Moloney, "Memory effects in the long-wave infrared avalanche ionization of gases: A review of recent progress," ArXiv:1810.07345 Phys (2018).

[10] T. Popmintchev, M-C. Chen, D. Popmintchev, P. Arpin, S. Brown, S. Alisauskas, *et al.*, "Bright coherent ultrahigh harmonics in the keV X-ray regime from mid-infrared femtosecond lasers," *Science* **336**, 1287–1291 (2012).

[11] F. Silva, S. M. Teichmann, S. L. Cousin, M. Hemmer, and J. Biegert, "Spatiotemporal isolation of attosecond soft X-ray pulses in the water window," *Nat. Commun.* **6**, 6611 (2015).

[12] J. C. Travers, W. Chang, J. Nold, N. Y. Joly, and P. St.J. Russell, "Ultrafast nonlinear optics in gas-filled hollow-core photonic crystal fibers," *J. Opt. Soc. Am. B* **28**, A11–26 (2011).

[13] P. Hölzer, W. Chang, J. C. Travers, A. Nazarkin, J. Nold, N. Y. Joly, M. F. Saleh, F. Biancalana, P. St.J. Russell, "Femtosecond nonlinear fiber optics in the ionization regime," *Phys. Rev. Lett.* **107**, 203901 (2011).

[14] E. E. Serebryannikov and A. M. Zheltikov, "Ionization-induced effects in the soliton dynamics of high-peak-power femtosecond pulses in hollow photonic-crystal fibers," *Phys. Rev. A* **76**, 013820 (2007).

[15] F. Köttig, D. Novoa, F. Tani, M. C. Günendi, M. Cassataro, J. C. Travers, *et al.*, "Mid-infrared dispersive wave generation in gas-filled photonic crystal fiber by transient ionization-driven changes in dispersion," *Nat. Commun.* **8**, 813 (2017).

[16] F. Köttig, F. Tani, J. C. Travers, and P. St.J. Russell, "PHz-wide spectral interference through coherent plasma-induced fission of higher-order solitons," *Phys. Rev. Lett.* **118**, 263902 (2017).

[17] F. Köttig, F. Tani, C. M. S. Biersach, J. C. Travers, and P. St.J. Russell, "Generation of microjoule pulses in the deep ultraviolet at megahertz repetition rates," *Optica* **4**, 1272–1276 (2017).

[18] T. K. Allison, A. Cingöz, D. C. Yost, and J. Ye, "Extreme nonlinear optics in a femtosecond enhancement cavity," *Phys. Rev. Lett.* **107**, 183903 (2011).

[19] J. R. Koehler, F. Köttig, B. M. Trabold, F. Tani, and P. St.J. Russell, "Long-lived refractive-index changes induced by femtosecond ionization in gas-filled single-ring photonic-crystal fibers," *Phys. Rev. Appl.* **10**, 064020 (2018).

[20] B. M. Trabold, M. I. Suresh, J. R. Koehler, M. H. Frosz, F. Tani, P. St.J. Russell, "Spatio-temporal measurement of ionization-induced modal index changes in a gas-filled PCF by prism-assisted side-coupling," *Opt. Exp.* **27**, 14392 (2019).

[21] M. Erkintalo, Y. Q. Xu, S. G. Murdoch, J. M. Dudley, and G. Genty, "Cascaded phase matching and nonlinear symmetry breaking in fiber frequency combs," *Phys. Rev. Lett.* **109**, 223904 (2012).

[22] X. Liu, A. S. Svane, J. Lægsgaard, J. Tu, S. A. Boppart, and Turchinovich D, "Progress in Cherenkov femtosecond fiber lasers," *J. Phys. Appl. Phys.* **49**, 023001 (2016).

[23] Y-H. Cheng, J. K. Wahlstrand, N. Jhajj, and H. M. Milchberg, "The effect of long timescale gas dynamics on femtosecond filamentation," *Opt. Exp.* **21**, 4740–4751 (2013).

[24] R. W. Boyd, *Nonlinear Optics*, (Third Edition, Academic Press, 2008).

[25] M. Kolesik, J. V. Moloney, and M. Mlejnek, "Unidirectional optical pulse propagation equation," *Phys. Rev. Lett.* **89**, 283902 (2002).

[26] D. P. Shelton Nonlinear-optical susceptibilities of gases measured at 1064 and 1319nm, *Phys. Rev. A* **42**, 2578 (1990).

[27] A. Couairon, and A. Mysyrowicz, "Femtosecond filamentation in transparent media," *Phys. Rep.* **441**, 47–189 (2007).

[28] See Supplemental Material at [URL will be inserted by publisher] for a simplified schematic of the procedure for obtaining the reconstruction of the plasma profile.

[29] X. Liu, J. Laegsgaard, R. Iegorov, A. S. Svane, F. Ö. Ilday, H. Tu, S. A. Boppart, and D. Turchinovich, "Nonlinearity-tailored fiber laser technology for low-noise, ultra-wideband tunable femtosecond light generation," *Photon. Res.* **5**, 750–761 (2007).